\documentclass[11pt]{article} 

\newtheorem{theorem}{Theorem}[section]
\newtheorem{lemma}[theorem]{Lemma}

\def\square{\rule{2mm}{2mm}}

\newenvironment{proof}{{\noindent\bf Proof:  }}{\qquad\square}

\def\squarebox#1{\hbox to #1{\hfill\vbox to #1{\vfill}}}

\newcommand{\tensor}{\otimes}
\newcommand{\Tensor}{\bigotimes}
\newcommand{\xor}{\oplus}

\newcommand\meet\wedge
\newcommand\implies\Rightarrow

\newcommand{\reals}{{\hbox{\sf I\kern-.14em\hbox{R}}}}
\newcommand{\trace}{{\rm Tr}}
\newcommand{\prob}{{\rm Pr}}
\newcommand{\size}[1]{\left|#1\right|}

\newcommand{\ket}[1]{|#1\rangle}
\newcommand{\bra}[1]{\langle #1|}
\newcommand{\braket}[2]{\langle #1 | #2\rangle}
\newcommand{\ketbra}[2]{\ket{#1}\!\bra{#2}}
\newcommand{\density}[1]{\ketbra{#1}{#1}}
\newcommand{\norm}[1]{\left\|\,#1\,\right\|}
\newcommand{\trnorm}[1]{\norm{#1}_{\mathrm {tr}}}
\newcommand{\set}[1]{{\left\{#1\right\}}}

\newcommand{\ignore}[1]{}

\newcommand{\inr}{\in_{\mathrm R}}
\newcommand{\diag}{{\mathrm{diag}}}

\newcommand{\cee}{{\mathrm c}}
\newcommand{\pee}{{\mathcal P}}

\begin{document}

\title{On bit-commitment based quantum coin flipping}

\author{
Ashwin Nayak \thanks{ Computer Science Department, and Institute for
  Quantum Information, California Institute of Technology, Mail
  Code~256-80, Pasadena, CA~91125, USA. Email: {\tt nayak@cs.caltech.edu}.
  Supported by Charles Lee Powell Foundation, and NSF grants
  CCR~0049092 and EIA~0086038. Part of this work was done while this
  author was at DIMACS Center, Rutgers University, and AT{\&}T
  Labs, and was supported by NSF grants STC~91-19999, CCR~99-06105 and
  EIA~00-80234.} \\
Caltech
\and
Peter Shor \thanks{ AT\&T Labs--Research, 180 Park Ave, Florham Park,
  NJ~07932, USA. Email: {\tt shor@research.att.com}.} \\
AT\&T Labs
}

\date{}

\maketitle

\begin{abstract}
In this paper, we focus on a special framework for quantum coin
flipping protocols, {\em bit-commitment based protocols}, within which  
almost all known protocols fit. We show a lower bound of~$1/16$ for the
bias in any such protocol.  We also analyse a sequence of multi-round protocol
that tries to overcome the drawbacks of the previously proposed
protocols, in order to lower the bias. We show an intricate cheating
strategy for this sequence, which leads to a bias of~$1/4$. This
indicates that a bias of~$1/4$ might be optimal in such protocols, and
also demonstrates that a cleverer proof technique may be required to
show this optimality.
\end{abstract}

\section{Quantum coin flipping}
\label{sec-intro}

Coin flipping is the communication problem in which two distrustful
parties wish to agree on a common random bit, by ``talking over the
phone''~\cite{Blum81}. When the two parties follow a protocol
honestly, the bit they agree on is required to be $0$ or $1$ with
equal probability. Ideally, they would also like that if any
(dishonest) party deviates from the protocol, they do not agree on any
particular outcome with probability more than~$1/2$. It is known
that ideal coin flipping is impossible, in both, the classical and the
quantum setting~\cite{LoC98,Mayers97}. In fact, in any classical
protocol, one of the two parties can force the outcome of the protocol
to a value of her choice with probability
$1$. In~\cite{AharonovTVY00}, Aharonov, Ta-Shma, Vazirani, and Yao
showed that it is possible to design a {\em quantum\/} coin flipping
protocol in which no player can force the outcome of the protocol with
probability more than a constant~$1/2 +
\epsilon$, with bias a constant~$\epsilon < 1/2$. In other words, any
cheating player in such protocols is detected with constant
probability. Later, Ambainis~\cite{Ambainis01} gave an improved
protocol with bias at most~$1/4$.

Formally, a quantum coin flipping protocol with bias~$\epsilon$ is a
two-party communication game in the style of~\cite{Yao93}, in which
the players start with no inputs, and compute values~$c_A, c_B \in
\set{0,1}$ respectively (or declare that the other player is
cheating). The protocol satisfies the following additional properties:
\begin{enumerate}
  
\item If both players are honest (i.e., follow the protocol), then
  they agree on the outcome of the protocol: $c_A = c_B$, and the
  outcome is $0$ or $1$ with equal probability: $\Pr(c_A = c_B = b) =
  1/2$, for $b \in \set{0,1}$.
  
\item If one of the players is honest (i.e., the other player may
  deviate arbitrarily from the protocol in his or her local
  computation), then the outcome of the protocol has bias at
  most~$\epsilon$: for any $b \in \set{0,1}$, $\Pr(c_A = c_B = b) \le
  1/2+ \epsilon$.
\end{enumerate}


Almost all quantum coin flipping protocols with bias that is provably smaller
than a half~\cite{AharonovTVY00,Ambainis01} are based on the notion of
bit-commitment. In other words, they have the following form, when the
parties flipping the coin, Alice and Bob, are honest.\footnote{We were
recently informed~\cite{Salvail01} of a protocol of a different kind that
also achieves a bias of~$1/4$.}

{\bf Protocol schema~$\Pi$}:
\begin{enumerate}
  
\item First, Alice and Bob each pick a random bit, $a$ and~$b$
  respectively, and privately construct states~$\rho_a$ and
  $\sigma_{b}$. The states are over three sets of qubits, a {\em
    commitment\/} part, a {\em revelation\/} part, and a {\em
    verification\/} part. The revelation part consists of one qubit
  that contains the value of the bit picked.
  
\item Next, they {\em commit\/} to their respective bits~$a$ and~$b$,
  by sending each other the commitment part of their states~$\rho_a$
  and~$\sigma_{b}$. They may do this over several rounds of
  communication, in which they send messages alternately.
  
\item Then, they {\em reveal\/} to the other party the bits~$a,b$
  they picked (in some order), and follow that up by sending the
  rest of the states~$\rho_a$,
  $\sigma_{b}$ (the verification part). This may again be
  over several rounds of communication. This allows 
  the each party to check via suitable measurements that
  the state with which the other, Alice (or Bob), committed to her
  (his) bit is indeed consistent with~$a$ ($b$).

\end{enumerate}
The result of the protocol is~$c = a \xor b$, if neither player
is detected cheating during the third (verification) stage.

For example, in the case of the protocol in~\cite{Ambainis01}, Alice
uses the right half of the following state to commit to her bit~$a$, and
the left half to help Bob check her commitment:
\begin{eqnarray*}
\rho_a & = &  \sum_{s = 0,1} \frac{1}{2} \density{a,s} \tensor
              \density{\psi_{a,s}}, \textrm{~~~~where} \\
\ket{\psi_{x,s}} & = & \frac{1}{\sqrt{2}} (\ket{0} + (-1)^s \ket{x+1}).
\end{eqnarray*}
Bob skips the commit stage, and directly reveals his bit~$b$. In the
final stage, Bob checks that the state of the qutrit Alice sent in the 
first round is indeed consistent with~$a,s$, by
measuring it in a basis containing~$\ket{\psi_{a,s}}$.

Protocols of the form described above may be recast in the following
terms.  First, Alice and Bob each pick a random bit.  Then, they
successively send each other qubits {\em which do not depend on the
  qubits sent by the other party in the previous rounds\/}. The qubits
sent by Alice represent a commitment to her bit~$a$ along with
auxiliary information required by Bob to check if she is cheating.
Similarly for the qubits sent by Bob. Thus, after all communication is
over, the states Alice and Bob sent to each other for~$0$ and~$1$ are
perfectly distinguishable. They measure the states received from the
other party (possibly with some ancilla) according to a von Neumann
measurement to determine the bits~$a,b$ or to detect cheating. The
outcome of the coin flip is~$a \xor b$ if no cheating is detected.  In
this description, we have assumed, w.l.o.g.~\cite{BernsteinV97}, that
all measurements are done at the end. Note that the description is
also slightly more general in that the players may not explicitly
reveal the bits they intend to commit to, and the commitment and the
verification stages may be interleaved. We will henceforth refer to
such protocols as {\em bit-commitment based protocols\/}.

In this paper, we study coin flipping protocols that fall into the
special framework described above, that of bit-commitment based
protocols.  We show a lower bound of~$1/16$ for the bias in any such
protocol (Theorem~\ref{thm-lb}).
This provides a single proof that these protocols,
including the one proposed in~\cite{MayersSC99}, cannot lead to
arbitrarily small bias.

Next, we analyse a sequence of protocols that tries to overcome the
drawbacks of the previously proposed protocols, and also tries to
circumvent the cheating strategy that leads to this above lower bound.
We show an intricate cheating strategy for the sequence of protocols,
which leads to a bias of~$1/4$.  This indicates that a bias of~$1/4$
might be optimal in such protocols, and also demonstrates that a
cleverer proof technique than the one used in Theorem~\ref{thm-lb} is
required to show this optimality.

Recently, Kitaev has shown a lower bound of~$1/\sqrt{2} - 1/2 =
0.2071\ldots$ for the bias in {\em arbitrary\/} quantum coin flipping
protocols~\cite{Kitaev01}. This is bigger than the bias of~$1/16 =
0.0625$ that we show (and applies to {\em any\/} coin-flipping
protocol). Kitaev's lower bound, however, doesn't seem to apply to
quantum games in which the two parties involved compete to ``win'' by
getting a particular value of the coin as the outcome (say Alice wins
if the outcome is~$0$, and Bob wins if the outcome is~$1$). This is
also known as {\em weak\/} coin-flipping in the literature. Protocols
for weak coin-flipping with bias less than~$1/4$ have been discovered.
Kerenidis and Nayak~\cite{KerenidisN01} have shown a protocol with
cheating probability at most~$0.739\ldots$.
Ambainis~\cite{Ambainis01b}, and Spekkens and
Rudolph~\cite{SpekkensR02} have shown protocols with an even lower
cheating probability of at most~$1/\sqrt{2} = 0.707\ldots$.

\section{A lower bound on the bias}

We first show that any bit-commitment based protocol may be reduced to
an extremely simple protocol of the same type, with bias at most
that in the original protocol,
and by increasing the number of rounds by at most~$1$.

\begin{lemma}
\label{thm-bcprotocol}
For any bit-commitment based coin flipping protocol~$\pee$ (of the
form~$\Pi$, or more generally, as described in
Section~\ref{sec-intro}), there is another such protocol~$\pee'$ such
that
\begin{enumerate}
  
\item The states~$\rho_a, \sigma_b$ are pure: $\rho_a =
  \density{\psi_a}$ and $\sigma_b = \density{\phi_{b}}$,
  where~$\ket{\psi_0} \perp \ket{\psi_1}$ and $\ket{\phi_0} \perp
  \ket{\phi_1}$,
  
\item Alice measures the state she received from Bob according to the
  measurement given by the operators~$P_0 = \density{\phi_0}$, $P_1 =
  \density{\phi_1}$, and $P_\cee = I - P_0 - P_1$, to determine Bob's
  bit or to check if he is cheating. Bob does an analogous measurement
  given by~$Q_0,Q_1,Q_\cee$ on the state he receives from Alice, and

\item The bias is at most the bias of~$\pee$.
\end{enumerate}
\end{lemma}
\begin{proof}
The protocol~$\pee'$ is obtained by stipulating that the players use a
fixed {\em purification\/} $\ket{\psi_a}, \ket{\phi_{b}}$ of the
states~$\rho_a, \sigma_{b}$ used in the original
protocol~$\pee$. Since the states $\rho_0, \rho_1$ are perfectly
distinguishable, their purifications are orthogonal. Similarly with
$\phi_0, \phi_1$. All but the last two rounds of~$\pee'$ are as in the
original one. We stipulate that the players send the entire (purified)
state $\psi_a$ or $\phi_{b}$ in~$\pee'$. Thus, the last player to send
a message in~$\pee$ sends the qubits used in the purification in the
penultimate round of~$\pee'$. In the final round, the other player
sends the qubits used in purifying her state. We also alter the
measurement to the ones mentioned in the lemma.

We now show that this modification of the protocol results in bias at
most that in the original one. We do this by showing that any cheating
strategy of a player in the modified protocol~$\pee'$ that achieves a
bias of~$\epsilon$ leads to a cheating strategy in the original
protocol~$\pee$ with at least the same bias.

For concreteness, we consider a cheating strategy for Alice in the
protocol~$\pee'$. (The argument for the case of Bob is similar.) In her
strategy in~$\pee$, Alice acts exactly as in the original strategy,
except that she is not required to send the ``purification qubits''
meant for her last message in~$\pee'$. We need only show that the
probability with which the measurement in~$\pee$ yields $0$ or $1$
dominates the same probability for~$\pee'$.

Suppose that Bob uses a von Neumann measurement given by the
projection operators~$R_0, R_1, R_\cee$ in~$\pee$. We concentrate on
the probability that Alice can convince Bob that she had picked~$a =
0$.  The other probability may be bounded similarly. Since~$R_0\rho_0
R_0 = \rho_0$, the purification~$\ket{\psi_0}$ lies in the range
of~$I\tensor R_0$. The states of the qubits sent to Bob by Alice and
her private qubits in~$\pee'$ are together given by some mixed
state~$\sum_j \mu_j \density{\xi_j}$ (where the states~$\ket{\xi_j}$
are over the space of~$\rho_a$ and the purification space). It thus
suffices to show that for any state~$\ket{\xi} \in \set{\ket{\xi_j}}$,
$$
\norm{I \tensor R_0 \ket{\xi}}^2 ~~\ge~~ \size{\braket{\psi_0}{\xi}}^2.
$$
Note that the LHS is the squared-norm of the projection
of~$\ket{\xi}$ onto the range of~$I\tensor R_0$, and the RHS is the
squared-norm of the projection of the same vector onto a {\em
  subspace\/} of that range, the one-dimensional space spanned
by~$\ket{\psi_0}$.  The inequality is then immediate. This shows that
Alice can achieve at least the same bias as in $\pee'$.
\end{proof}

This simple characterisation of bit-commitment based protocols proves
useful in the analysis of the smallest bias achievable with such
protocols. Using this, we show that coin flipping protocols based on
bit-commitment cannot achieve arbitrarily small bias.

\begin{theorem}
\label{thm-lb}
In any quantum coin flipping protocol based on bit-commitment, one of
the parties can achieve probability of cheating at least~$9/16$.
\end{theorem}
\begin{proof}
As shown in Lemma~\ref{thm-bcprotocol}, any such protocol between honest
parties may be viewed as follows: first, Alice and Bob construct the
states~$\ket{\psi_a}$ and~$\ket{\phi_{b}}$, respectively,
corresponding to the random bits~$a$ and~$b$.  Then, they send each
other a part of the states~$\ket{\psi_a}, \ket{\phi_{b}}$ a few
qubits at a time. Finally, they measure the qubits received from each
other using projections~$P_0, P_1, P_{\mathrm c}$ and~$Q_0, Q_1,
Q_{\mathrm c}$.
  
Let~$\rho_{a,i} = \trace_{A_i} (\density{\psi_a})$ be the state sent
to Bob by Alice by round~$i$ (so~$A_i$ are the qubits
of~$\ket{\psi_a}$ still with Alice after the~$i$-th round).
Let~$\sigma_{b,i}$ be the corresponding state sent to Alice by Bob.
  
Let there be~$n$ rounds in all.  Let~$F_{A,i} = F(\rho_{0,i},
\rho_{1,i})$ and similarly~$F_{B,i} = F(\sigma_{0,i},
\sigma_{1,i})$. Here,~$F(\cdot,\cdot)$ is {\em fidelity\/} function as
defined in~\cite{Jozsa94}.  So~$F_{A,0}
= F_{B,0} = 1$.  Note that~$Q_a \rho_{a,n} Q_a = \rho_{a,n}$, and
similarly~$P_{b} \sigma_{b,n} P_{b} = \sigma_{b,n}$, so that~$F_{A,n}
= F_{B,n} = 0$.

\begin{lemma}
\label{thm-alpha}
Consider a protocol with honest players.  For any constant~$0 \le
\alpha \le 1$, there is a player, say Alice, and a round~$k \ge 0$
such that the states she sends to Bob by the~$k$-th round on~$a=0$
and~$1$ have fidelity at least~$\alpha$, and the fidelity of the
states she receives from Bob by the next round have fidelity at
most~$\alpha$. In other words, $F_{A,k} \ge \alpha$ and $F_{B,k+1} \le
\alpha$.
\end{lemma}
\begin{proof}  
The case~$\alpha = 1$ is trivial. Let~$\alpha < 1$.  Note that
both~$F_{A,i},F_{B,i}$ decrease from~$1$ to~$0$ through the course of
the protocol.  Consider the first round~$i \ge 1$ such that one of
these, say~$F_{B,i}$, becomes~$\le \alpha$.  Round~$k = i-1$ satisfies
the property we seek.
\end{proof}

We will devise a cheating strategy for a player as given by
Lemma~\ref{thm-alpha} above for~$\alpha = 1/4$. Say this player is
Alice, and the round identified in the lemma is~$k$. There is a
unitary transformation on the qubits~$A_k$ which achieves maximum fidelity
between~$\rho_{0,k},\rho_{1,k}$~\cite{Jozsa94}, i.e., $$
\size{\bra{\psi_0}U\ket{\psi_1}}^2 = F_{A,k} = F(\rho_{0,k},
\rho_{1,k}) \ge {1\over 4}.
$$
Moreover, we may assume that~$\bra{\psi_0}U\ket{\psi_1}$ is real and
non-negative.

Alice may cheat as follows. She constructs the state
$$
\ket{\xi} = \frac{\ket{\psi_0} + U\ket{\psi_1}}{\norm{\psi_0 + U\psi_1}}
$$
and uses this state in the protocol till round~$k+1$. After this round,
she makes the best possible measurement to distinguish~$\sigma_{0,k+1}$
and~$\sigma_{1,k+1}$ to guess the value of~$b$. If her guess~$g$ is~$0$,
she proceeds with the rest of the protocol, Otherwise, she
applies~$U^\dag$ to her part of~$\ket{\xi}$ (the qubits in~$A_k$) 
and then completes the protocol.

\begin{lemma}
$\prob(c = 0|g = b) \ge \frac{1 + \sqrt{F_{A,k}}}{2}$.
\end{lemma}
\begin{proof}
First, note that
$$
\norm{\psi_0 + U \psi_1}^2 ~~=~~ \norm{\psi_0}^2 + \norm{\psi_1}^2 + 2
\bra{\psi_0}U\ket{\psi_1} ~~=~~ 2(1 + \sqrt{F_{A,k}}).
$$

Suppose~$b = 0$. (The other case is similar.)  Note that the
probability that Alice succeeds in getting the outcome~$c = 0$ given
that she guesses the value of~$b$ correctly is
$$
\norm{Q_0 \xi}^2 ~~=~~ \frac{\norm{Q_0 \psi_0 + Q_0 U \psi_1}^2}{
  \norm{\psi_0 + U\psi_1}^2 } ~~=~~ \frac{ \norm{Q_0\psi_0}^2 +
\norm{Q_0
    U \psi_1}^2 + 2 \bra{\psi_0} Q_0 U \ket{\psi_1} }{ 2 (1 +
  \sqrt{F_{A,k}}) }.
$$
Now~$Q_0\ket{\psi_0} = \ket{\psi_0}$, since~$\ket{\psi_0}$ is the
state Alice would have used if she were honest. So the first term in the
numerator above is~$1$, and the last term is~$2 \sqrt{F_{A,k}}$. The
second term may be bounded from below by noting that
since~$\ket{\psi_0}$ belongs to the range of~$Q_0$, the square-norm of
the projection~$Q_0 U\ket{\psi_1}$ is at
least~$\bra{\psi_0}U\ket{\psi_1}^2 = F_{A,k}$. Thus, the probability
of cheating is at least
$$
\frac{1 + F_{A,k} + 2 \sqrt{F_{A,k}}}{2(1 + \sqrt{F_{A,k}})} ~~=~~
\frac{1 + \sqrt{F_{A,k}}}{2},
$$
which is the bound claimed.
\end{proof}

The probability that Alice correctly guesses~$b$ is (using Bayes'
strategy)
$$
\prob(g = b) ~~=~~ {1\over 2} + \frac{\trnorm{\sigma_{0,k+1} -
\sigma_{1,k+1}}}{4}.
$$
By a result of Fuchs and van de Graaf~\cite{FuchsG99},
$$
\trnorm{\sigma_{0,k+1} - \sigma_{1,k+1}} ~~\ge~~ 2(1 -
\sqrt{F_{B,k+1}}) ~~\ge~~ 2(1 - {1\over 2}) ~~=~~ 1.
$$
The net probability that Alice succeeds in biasing the coin towards~$0$
is therefore
$$
\prob(c = 0) ~~\ge~~ \prob(c = 0|g = b) \cdot \prob(g = b) 
~~\ge~~ {3 \over 4}\cdot {3 \over 4} ~~=~~ {9\over 16}.
$$
This proves the theorem.
\end{proof}

\section{A sequence of highly interactive protocols}

In this section we look at a sequence of bit-commitment based
protocols in which Alice and Bob very gradually send each other
information about their bits in the commit stage. Intuitively, such
protocols seem to be good candidates for achieving bias much smaller
than~$1/4$, since a dishonest player does not get much information
about the other's bit until a significant number of rounds have
elapsed, and he would have heavily committed to some bit by then.
However, this intuition appears to be mistaken, and we give intricate
cheating strategies for each of the players with which at least one of
them can achieve bias at least as high as~$1/4$. This suggests that
the optimal bias for this kind of protocol might be~$1/4$, and also
that proving this optimality might require ideas more sophisticated
than those in Theorem~\ref{thm-lb}.

Define, for~$x,s \in \set{0,1}$,
$$
\ket{\psi(x,s)} ~~=~~
\sqrt{1-\epsilon}\,\ket{0} + (-1)^{s} \sqrt{\epsilon}\,\ket{x+1}.
$$ 
These states provide the best trade-off between how much
information they reveal, and how much cheating in commitment they
allow.

The protocol~$\pee_n$, $n \in \set{1,2,3,\ldots}$, goes as follows.
Alice picks~$a \inr \set{0,1}$, and Bob picks~$b \inr \set{0,1}$.
Then they alternately send each other, for a total
of~$n$ rounds, the states~$\ket{\psi(a,s)}$ and~$\ket{\psi(b,s)}$
respectively, for independently chosen random sign~$s$,
starting with Alice. The last player to receive such a state then 
reveals the bit he/she chose, followed by the other
player. Then, they reveal the signs used in their states in the 
opposite order, and the other party checks the state he/she
received against the
claimed bit and sign. If no cheating is detected, the players 
declare the~$c = a\xor b$ as the result of the protocol.

More formally,
\begin{enumerate}
\item 
For~$i = 1, 2, 3, \ldots, n$, 
if~$i$ is odd, Alice picks~$s_i \inr \set{0,1}$, and 
sends~$\ket{\psi(a,s_i)}$ to Bob, else, if~$i$ is even, Bob
picks~$s_i \inr \set{0,1}$, and sends~$\ket{\psi(b,s_i)}$ to Alice.

\item
If~$n$ is even, Alice sends~$a$ to Bob, and then Bob sends~$b$ to
Alice. Otherwise, if~$n$ is even they reveal their bits in the
opposite order.

\item
For~$i = n, n-1, n-2, \ldots, 1$, the player that picked~$s_i$ reveals
it to the other player.  In other words, the ``signs'' used in the
states are revealed in the opposite order: If~$i$ is odd, Alice
sends~$s_i$ to Bob, else Bob sends~$s_i$ to Alice. The player that
receives this bit checks via a measurement that the state sent to
her/him in the~$i$-th round of the protocol is indeed consistent with
the bit and the sign that the other player sent.

Note that the order of revealing signs is designed so that the
na\"{\i}ve strategy of reusing a state that a player got in a previous
round does not work.

\item
If {\em all\/} the checks are passed, the outcome of the protocol
is~$c = a \xor b$.
\end{enumerate}

Before we give cheating strategies for these protocols, we analyse
general versions of~$\pee_1$ and~$\pee_2$. This illustrates the main
approach taken in the strategies for~$\pee_n$, $n\ge 3$.

\subsection{The three-round version}
\label{sec-three}

We start by analysing the protocol~$\pee_1$ with one round of
commitment. This happens to be a parametrised version of the
three-round protocol due to Ambainis~\cite{Ambainis01}. We prove a
property of this protocol that helps us analyse protocols with more
rounds.

The protocol may be described with a parameter~$\alpha \in [0,\pi]$
(such that~$\tan{\alpha\over 2} =
\sqrt{\frac{\epsilon}{1-\epsilon}}\;$) as:
\begin{enumerate}
\item Alice picks~$a,s \inr \set{0,1}$, and sends Bob the
      state~$\ket{\psi_{a,s}}$, where the state is defined as follows:
\begin{eqnarray}
\ket{\psi_{a,s}}
    & = & \cos{\alpha\over 2}\ket{0}
          + (-1)^s \sin{\alpha\over 2}\ket{a+1}.
\end{eqnarray}

\item Bob picks~$b\inr \set{0,1}$ and sends it to Alice.

\item Alice then reveals the bits~$a,s$ to Bob, who checks for
consistency with the state initially sent by Alice.
\end{enumerate}
The output of the protocol is given by~$c = a \xor b$, if all the
checks are passed.

\begin{lemma}
\label{thm-three-round}
If Bob is honest, then~$\prob(c = 0) \le {1 \over 4}(3 + \cos\alpha)$.
\end{lemma}
\begin{proof}
The analysis proceeds as in~\cite{Ambainis01}, by symmetrising the
strategy of a dishonest Alice,
so that in the last round, she sends~$s = 0$ and~$s = 1$ with 
equal probability, and assuming that she always sends~$a = b$
(this only increases her chances of cheating successfully). For such a
symmetric strategy, Ambainis~\cite[Lemma~8]{Ambainis01} shows
that
\begin{eqnarray}
\label{eqn-prob-bound}
\prob(c = 0) & \le & \frac{F(\rho_0,\rho) + F(\rho_1,\rho)}{2},
\end{eqnarray}
where~$\rho$ is the state of Bob after the first round, and~$\rho_0,
\rho_1$ are the analogous states corresponding to~$b = 0,1$
respectively, if Alice were honest:
\begin{eqnarray*}
\rho_a & = & {1\over 2} ( \density{\psi_{a,0}} + \density{\psi_{a,1}}),
\end{eqnarray*}
and~$F(\sigma_0,\sigma_1) = \trnorm{\sqrt{\sigma_0}\sqrt{\sigma_1}}^2$
denotes the fidelity of two density matrices~\cite{Jozsa94}.

We show in Lemma~\ref{thm-fidelity} below that the expression in
equation~(\ref{eqn-prob-bound}) is bounded above by 
$$ 
{1\over 2} (1 + F(\rho_0,\rho_1)^{1/2}). 
$$ 
Since~$F(\rho_0,\rho_1) = \cos^4 {\alpha\over 2}$, the bound follows.
\end{proof}

We now prove the lemma mentioned above.
\begin{lemma}
\label{thm-fidelity}
For any two density matrices~$\sigma_0,\sigma_1$,
$$
\max_\sigma ~~ F(\sigma_0,\sigma) + F(\sigma_1,\sigma)
    ~~\le~~ 1 + F(\sigma_0,\sigma_1)^{1/2}.
$$
\end{lemma}
\begin{proof}
Let~$\sigma$ be the density matrix that achieves the maximum, and
let~$\ket{\phi_x}$ be a purification of~$\sigma_x$, for~$x = 0,1$.
Let~$\ket{\xi_x}$ be the purification of~$\sigma$ that achieves maximum
fidelity with~$\sigma_x$~\cite{Jozsa94}: $$ F(\sigma_0,\sigma) ~~=~~
\size{\braket{\phi_x}{\xi_x}}^2. $$ Since~$\ket{\xi_0},\ket{\xi_1}$
are the purifications of the same density matrix~$\sigma$, there is a
local unitary operator~$U$ such that~$\ket{\xi_1} = U \ket{\xi_0}$.
Now,
\begin{eqnarray}
F(\sigma_0,\sigma) + F(\sigma_1,\sigma)
    & = & \size{\braket{\phi_0}{\xi_0}}^2 +
          \size{\braket{\phi_1}{\xi_1}}^2  \nonumber \\
    & = & \size{\braket{\phi_0}{\xi_0}}^2 +
          \size{\bra{\phi_1}U\ket{\xi_0}}^2 \nonumber \\
    & \le & \max_{\ket{\xi}} \size{\braket{\phi_0}{\xi}}^2 +
            \size{\bra{\phi_1}U\ket{\xi}}^2 \nonumber \\
\label{eqn-eval}
    & = & 1 + \size{\bra{\phi_1}U\ket{\phi_0}} \\
    & \le & 1 + \max_{\mathrm{local}~U}
            \size{\bra{\phi_1}U\ket{\phi_0}} \nonumber \\
\label{eqn-fidelity}
    & = & 1 + F(\sigma_0,\sigma_1)^{1/2}.
\end{eqnarray}
Equation~(\ref{eqn-eval}) above follows by noticing that the
state~$\ket{\xi}$ that achieves the maximum is the vector that bisects
the angle between~$\ket{\phi_0}$ and~$U\ket{\phi_1}$. Another way of
getting the bound is by noticing that the expression is the maximum
eigenvalue of the matrix~$\density{\phi_0} +
U\density{\phi_1}U^\dag$. The last
step---equation~(\ref{eqn-fidelity})---follows from a characterisation
of fidelity due to Jozsa~\cite{Jozsa94}.
\end{proof}

\subsection{The five-round protocol}

Next, we give a tight analysis for a general five-round version of the
protocol~$\pee_2$ described above. This version of the protocol still
does not improve over the bias of~$1/4$ achieved by the three-round
protocol of~\cite{Ambainis01}. However, it suggests a better cheating
strategy for the many-rounds version than the one given in the proof
of Theorem~\ref{thm-lb}.

The version of the protocol~$\pee_2$ we consider has the following five
rounds with honest players:
\begin{enumerate}
\item Alice picks~$a,s \inr \set{0,1}$, and sends Bob the
  state~$\ket{\psi_{a,s}}$, where the state is defined as follows:
\begin{eqnarray}
\label{eqn-alice-state}
\ket{\psi_{a,s}}
    & = & \cos{\alpha\over 2}\ket{0}
          + (-1)^s \sin{\alpha\over 2}\ket{a+1},
\end{eqnarray}
for an angle~$\alpha$ to be specified later.

\item Similarly, Bob picks~$b,s' \inr \set{0,1}$ and sends Alice the
  state~$\ket{\phi_{b,s'}}$, where the state is defined as follows:
\begin{eqnarray}
\label{eqn-bob-state}
\ket{\phi_{b,s'}}
    & = & \cos{\beta\over 2}\ket{0}
          + (-1)^{s'} \sin{\beta\over 2}\ket{b+1},
\end{eqnarray}
for an angle~$\beta$ to be specified later.

\item Alice then reveals the bit~$a$ to Bob.
  
\item Bob reveals both~$b$ and~$s'$ to Alice, and Alice verifies, by
  an appropriate measurement, that the state sent by Bob is consistent
  with~$b,s'$.
  
\item Alice now discloses~$s$ as well, and Bob verifies that the state
  sent by Alice in the first round is consistent with~$a,s$.
\end{enumerate}
The output of the protocol, the coin flip~$c$, is given by the
exclusive-or of two bits, $c = a \xor b$ (provided no cheating is
detected).

In the case the players are honest, $\prob(c = 0) = \prob(c = 1) =
1/2$.  Below we prove an upper bound on the probability that any
player can achieve (by deviating from the protocol) for an outcome of
their choice. In the following discussion, we will assume, w.l.o.g.,
that a dishonest player prefers the outcome~$c = 0$.

First, we prove a bound on the probability that a dishonest Bob can
achieve, provided Alice is honest.
\begin{lemma}
\label{thm-five-bob}
If Alice is honest, then~$\prob(c = 0) ~\le~ 1 - {1 \over
2}\cos^2\frac{\alpha}{2} \sin^2\frac{\beta}{2}$.
\end{lemma}
\begin{proof}  
We claim that Bob's optimal cheating strategy is to measure the state
received from Alice in the standard basis, and then commit to a state
according to the outcome. If he observes~$\ket{1}$ or~$\ket{2}$, he
can cheat with probability~$1$ in the rest of the protocol. In case he
observes~$\ket{0}$, his probability of cheating successfully is
bounded by a constant less than one. This gives us the bound, as
explained below.
  
Note that the last round is of no consequence to Bob's cheating
strategy, and we may trace the sign bit~$s$ out after Alice sends the
first message (and eliminate the last round).  The protocol then
becomes equivalent to one in which the first round takes the following
form:
\begin{itemize}
\item[] Alice picks~$b \inr \set{0,1}$, and sends~$\ket{0}$ with
  probability~$\cos^2{\alpha\over 2}$, and~$\ket{b+1}$ with
  probability~$\sin^2{\alpha\over 2}$.
\end{itemize}
This reduces the protocol to a convex combination of two protocols, with
weights~$\cos^2{\alpha\over 2}$ and~$\sin^2{\alpha\over 2}$. In the
first protocol, the first message is the same regardless of the value
of~$b$, and in the second, the first message reveals~$b$ entirely.

In the first case, the protocol reduces to a three-round protocol of
the type studied in Section~\ref{sec-three}, with the role of Alice
and Bob reversed.  We may now use Lemma~\ref{thm-three-round} to bound
the probability that Bob can cheat by~$(3+\cos\beta)/4$.

In the second case, the protocol becomes trivial, and Bob can cheat
with probability~$1$.

The probability of convincing Alice that~$c = 0$ is thus bounded by
$$ 
\frac{3+\cos\beta}{4} \cos^2{\alpha\over 2} ~+~ \sin^2{\alpha\over 2},
$$
which reduces to the bound we seek.
\end{proof}

Bob can easily achieve the probability bound stated in the lemma, by
following a cheating strategy as in~\cite[Lemma~10]{Ambainis01}. This
shows that the analysis in our proof is optimal.

We now turn to the case where Alice is dishonest. The following lemma
bounds Alice's cheating probability. 

\begin{lemma}
If Bob is honest, then~$
\Pr(c = 0) ~~\le~~
\frac{1}{2} (1 + \cos^2 \frac{\alpha}{2}
\sin^2\frac{\beta}{2}).
$
\end{lemma}
\begin{proof}
If Alice is dishonest, she tries to force the
outcome~$c = 0$ by guessing the bit that Bob picked,
and then convincing him that the state she sent is
consistent with that guess.

We begin by symmetrising Alice's strategy so that the density matrix
of the qutrit she commits to in the first round is diagonal in
the~$0,1,2$ basis.  This is done in a manner as in
in~\cite[Lemma~6]{Ambainis01}; we omit the details. We may
therefore assume, w.l.o.g., that the joint state of Alice and Bob
after the communication in the first round is
$$
\ket{\xi} ~~=~~ \sqrt{\lambda_0}\, \ket{0,0} +
\sqrt{\lambda_1}\, \ket{1,1} +
\sqrt{\lambda_2}\, \ket{2,2}, 
$$
where~$\sum_i \lambda_i = 1$ and the density matrix of Bob's part is
$$
\rho ~~=~~ \lambda_0 \, \density{0} + \lambda_1 \,
\density{1} + \lambda_2 \, \density{2}.
$$

Alice's strategy in the third round is given by some
measurement on her entangled qutrit, the qutrit sent
by Bob, and some ancilla. This measurement gives the
value of the
bit she sends to Bob in that round. We consider the
superoperator~$T$ acting on their joint state
consisting of this measurement, composed with the
tracing out of all subsystems except the bit sent in
that round, and the qutrit sent in the first round.
This superoperator is determined  entirely by its
action on~$\density{i} \tensor \density{\xi}$, for~$i
= 0,1,2$, since the qutrit sent by Bob is a mixture of
the states~$\density{i}$. Let
\begin{eqnarray*}
T(\density{i} \tensor \density{\xi}) 
  & = & \density{0} \tensor \rho_{i0} + \density{1} \tensor \rho_{i1},
\end{eqnarray*}
where~$\sum_j \rho_{ij} = \rho$.
Thus, the unnormalised density matrix of the qutrit
Alice sent in the first round is:
\begin{equation}
\label{eqn-rho}
\cos^2\frac{\beta}{2}\; \rho_{0b} +
\sin^2\frac{\beta}{2}\; \rho_{b+1,b},
\end{equation}
given that Bob had picked bit~$b$, and Alice's guessed it
correctly:~$a = b$.

Let~$\tilde{\rho}_{ij}$ be the state, diagonal in the~$0,1,2$ basis, 
obtained by measuring the state~$\rho_{ij}$, and let:
\begin{eqnarray*}
\tilde{\rho}_{00} & = & \diag(\mu_0,\mu_1,\lambda_2-\mu_2) \\
\tilde{\rho}_{01} & = & \diag(\lambda_0-\mu_0,\lambda_1-\mu_1,\mu_2),
\end{eqnarray*}
where~$0 \le \mu_i \le \lambda_i$.
Finally, let~$
\tilde{\rho}_b ~=~ \cos^2\frac{\beta}{2} \; \tilde{\rho}_{0b} ~+~
\sin^2\frac{\beta}{2} \; \tilde{\rho}_{b+1,b}$
be the state~(\ref{eqn-rho}) measured in the~$0,1,2$ basis.
Let~$\sigma_b$ be the density matrix of the qutrit
Alice would send in the first round, if she were
honest, and had picked the bit~$b$:
$$
\sigma_b ~~=~~ \cos^2 \frac{\alpha}{2} \; \density{0} +
\sin^2 \frac{\alpha}{2} \; \density{b+1}.
$$

Assume now that instead of having picked~$s'$ at random, Bob had
created a uniform superposition over the two possible values for the
sign bit, and that he sends the qubit containing~$s'$ in the fourth
round.  It is now not hard to prove (cf.~\cite{Ambainis01}) that the
probability that (the dishonest) Alice is able to convince Bob that
the state she sent is consistent with~$b$, is at most~$
F(\tilde{\rho}_b, \sigma_b) $.  This probability is thus also bounded 
by~$F(\tilde{\sigma}_b,\sigma_b)$, where
\begin{eqnarray*}
\tilde{\sigma}_0
  & = & \cos^2\frac{\beta}{2} \; (\mu_0 \, \density{0} + \lambda_1 \,
        \density{1}) ~+~ \sin^2\frac{\beta}{2} \; \tilde{\rho} \\
\tilde{\sigma}_1
  & = & \cos^2\frac{\beta}{2} \; ((\lambda_0 - \mu_0) \, \density{0} 
        + \lambda_2 \, \density{2}) 
        ~+~ \sin^2\frac{\beta}{2} \; \tilde{\rho}.
\end{eqnarray*}
In other words, it only helps Alice to claim that she sent a state
corresponding to bit~$b$ if she either sees~$b+1$ in the qutrit she
receives from Bob, or if she sees~$0$ and had sent the state~$\ket{b+1}$ to
Bob in the first round.

By optimising over the choice of~$\mu_0$ for fixed~$\lambda_i$ and 
then optimising~$\lambda_i$,
we see that the optimum of~$F(\tilde{\sigma}_0, \sigma_0) +
F(\tilde{\sigma}_1, \sigma_1)$ is achieved when
\begin{eqnarray*}
\mu_0 
  & = & \frac{\lambda_0 (\lambda_1 - \lambda_2 \sin^2 {\beta\over
         2})}{ (\lambda_1 + \lambda_2) \cos^2 {\beta\over 2} }, \\
\lambda_1
  & \ge & \lambda_2 \; \sin^2 {\beta\over 2}, \textrm{~~~and} \\
\lambda_2
  & \ge & \lambda_1 \; \sin^2 {\beta\over 2}.
\end{eqnarray*}
We thus get the following bound on Alice's cheating
probability:
$$
\frac{1}{2} (1 + \cos^2 \frac{\alpha}{2}
\sin^2\frac{\beta}{2}),
$$
which is the bound claimed.
\end{proof}

Next, we describe a cheating strategy for Alice that achieves the
outcome of her choice with probability as high as in the upper bound
above, showing that our analysis is optimal.
\begin{lemma}
If Bob is honest, Alice can achieve~$
\Pr(c = 0) 
    ~\ge~ \frac{1}{2} (1 + \cos^2 \frac{\alpha}{2}
\sin^2\frac{\beta}{2}).
$
\end{lemma}
\begin{proof}
Alice constructs the following entangled state and sends one half of
it to Bob in the first round:
\begin{eqnarray*}
\ket{\xi} & = & \sqrt{1-\lambda} \ket{0,0} 
                   + \sqrt{\frac{\lambda}{2}} \ket{1,1} 
                   + \sqrt{\frac{\lambda}{2}} \ket{2,2},
                \textrm{~~~where} \\
\lambda & = & \frac{\sin^2\frac{\alpha}{2}}
                   { (1 + \sin^2\frac{\beta}{2})
                        \cos^2 \frac{\alpha}{2} +
                     \sin^2\frac{\alpha}{2} },
\end{eqnarray*}
and so the density matrix of the qutrit with Bob is~$\rho = \diag(
1-\lambda, \lambda/2, \lambda/2)$ in the~$0,1,2$ basis.

After Alice receives a qutrit from Bob in the second round, she
applies a unitary transformation to ``guess'' a value for~$b$ to
maximise her chances of getting~$c = 0$. This transformation acts
on the qutrit she received from Bob,
her entangled qutrit from the first round, 
and an ancilla qubit. It can be written as:
$$
\density{0} \tensor ( \density{0} \tensor H + \density{1} \tensor I +
                      \density{2} \tensor \sigma_{\mathrm x} ) 
~~+~~ \density{1} \tensor I \tensor I
~~+~~ \density{2} \tensor I \tensor \sigma_{\mathrm x}.
$$
where~$H$ is the Hadamard transform, and~$\sigma_{\mathrm x}$ is the
Pauli ``bit flip'' matrix.
In other words, Alice guesses~$a = b$, if the qutrit from Bob reveals
the identity of~$b$. Otherwise, if she committed with a~$\ket{d}$,
($d = 1,2$) she says~$a = d-1$, since this commitment is
irrevokable. If she committed with~$\ket{0}$, she says~$a = 0,1$ with
equal amplitude. It is crucial that she does this {\em in superposition\/}.

In the fourth round, Bob reveals the state of the qutrit he sent. If
Alice sees that the bit Bob picked is different from her guess,
i.e.~$b \not= a$, then she sends, say, ~$s = 0$ (an arbitrary value,
since she has lost the game). Otherwise, she tries to pick the
best~$s$ possible to maximise her chance of passing Bob's check. We
describe her actions when~$b = 0, s' = 0$; the other cases are
similar.

The unnormalised density matrix of the qutrit she sent in the first
round, when~$b = 0, s' = 0$, conditioned on her reply being~$a =
0$ is:
$$
\cos^2\frac{\beta}{2} \left( \frac{1-\lambda}{2} \; \density{0} +
\frac{\lambda}{2} \; \density{1} \right) ~~+~~ \sin^2\frac{\beta}{2}\; \rho.
$$
Note that she knows the state of all their qubits,
given~$b,s',a$. She can thus transform her part of the state so that
their joint state looks like
$$
\sqrt{\frac{1-\lambda}{2} (1 + \sin^2\frac{\beta}{2})} \; \ket{00}
~~+~~ \sqrt{\frac{\lambda}{2}} \; \ket{11}
~~+~~ \sqrt{\frac{\lambda}{2}} \, \sin \frac{\beta}{2} \; \ket{22}.
$$
She sends~$s = 0$ if the entangled bit is~$2$. If her entangled
qutrit is not~$2$, she does a Hadamard transform on her entangled
qubit, and sends that to Bob as~$s$.  The probability with which Bob
accepts~$a = b, s$ is then
$$
\left( \sqrt{\frac{1-\lambda}{2} (1 + \sin^2\frac{\beta}{2})}\;
\cos\frac{\alpha}{2}
+ \sqrt{\frac{\lambda}{2}}\; \sin\frac{\alpha}{2} \right)^2,
$$
which evaluates to the expression stated in the lemma.  She can
achieve the same probability of success for all other values of~$b,s'$
as well. Thus, the overall chance of her succeeding in cheating is
also given by this expression.
\end{proof}

The properties we established above show that this protocol still has a
bias of~$1/4$:
the cheating probability for Alice and Bob are of the form~$\frac{1}{2}
+ \delta$ and~$1-\delta$ respectively (with $\delta =
\frac{1}{2} \cos^2\frac{\alpha}{2} \sin^2\frac{\beta}{2}$), and their
maximum is minimised when~$\delta = 1/4$.

\subsection{Cheating strategies for~$\pee_n$}

Let~$A_n(\epsilon), B_n(\epsilon)$ be the maximum value of Alice 
and Bob's cheating probability in the protocol~$\pee_n$,
when the other party is honest.

Bob's optimal strategy may be reduced to a strategy for Alice as in
the proof of Lemma~\ref{thm-five-bob}.  Alice always starts the
protocol, and also sends the last message. Since the last message does
not affect Bob's cheating strategy, we may trace it out of the
protocol when analysing his optimal strategy. The protocol~$\pee_n$
then reduces to a mixture of protocols where Alice sends~$\ket{0}$ in
the first round with probability~$1-\epsilon$ (i.e.\ does not reveal
any information about the bit~$a$), and sends~$\ket{a+1}$ with
probability~$\epsilon$ (i.e.\ completely reveals the bit~$a$). The
rest of the protocol is the same as~$\pee_{n-1}$ with the roles of
Alice and Bob reversed. Thus,
\begin{equation}
\label{eqn-bn0}
B_n(\epsilon)
    ~~=~~ \epsilon + (1-\epsilon) A_{n-1}(\epsilon),
           \textrm{~~~~for } n \ge 2.
\end{equation}
We already know from~\cite{Ambainis01} that~$B_1(\epsilon) =
(1+\epsilon)/2$.
It thus suffices to analyse Alice's cheating probability in all 
the protocols~$\pee_n$.

From our analysis of the three and five round protocols, we also
know that
\begin{eqnarray*}
A_1(\epsilon) & = & 1 - \frac{\epsilon}{2} \\
B_2(\epsilon) & = & 1 - \frac{\epsilon}{2} + \frac{\epsilon^2}{2} \\
A_2(\epsilon) & = & \frac{1}{2} + \frac{\epsilon}{2} -
\frac{\epsilon^2}{2}.
\end{eqnarray*}

We now give a lower bound for the 
probability~$A_n(\epsilon)$ by describing a cheating strategy for
Alice that generalises the strategy that we saw in the five round
protocol.

First, we assume~$n = 2k-1$ ($k \ge 1$) is odd, so that Alice has~$k$
rounds of commitment, and Bob has~$k-1$. The case of~$n$ even is
addressed later.  In the first round, Alice sends one half of the
state 
$$
\sqrt{1-\lambda_1}\, \ket{00} ~+~ \sqrt{\frac{\lambda_1}{2}}\, \ket{11}
~+~ \sqrt{\frac{\lambda_1}{2}}\, \ket{22}
$$
to Bob, and retains the other half. The half she keeps is referred to
as an ``entangled qutrit'' below. All the parameters~$\lambda_j$ will
be specified later.  If one of the entangled qutrits she retains from
any previous round is in state~$x+1$, in all the commitment
rounds that follow, she sends the right half of the state 
$$
\frac{1}{\sqrt{2}}\; \ket{0}\ket{\psi(x,0)} ~+~
\frac{1}{\sqrt{2}}\; \ket{1}\ket{\psi(x,1)}
$$
and keeps the first qubit (called a ``sign qubit'' below). If the
entangled qutrits are all~$0$, and at least one of the qutrits she
received in earlier rounds from Bob is in state~$b+1$, then she sends
the right half of the state 
$$
\frac{1}{\sqrt{2}}\; \ket{0}\ket{\psi(b,0)} ~+~
\frac{1}{\sqrt{2}}\; \ket{1}\ket{\psi(b,1)}
$$
to Bob. 
Otherwise, if none of the above two events occurs, she sends one half
of the state
$$
\sqrt{1-\lambda_j}\; \ket{00} ~+~ \sqrt{\frac{\lambda_j}{2}}\; \ket{11}
~+~ \sqrt{\frac{\lambda_j}{2}}\; \ket{22}
$$
in the~$j$-th commitment round. It is important that she do all this
``in superposition,'' i.e., via unitary operations controlled by her
entangled qutrits and the qutrits she receives from Bob. She performs no
measurements in the process.

Formally, for each qutrit that Alice is supposed to send, she has a
qutrit-qubit pair. The first serves as the ``entangled qutrit'', and
the second serves as the ``sign qubit''. They are all initialised to
the~$0$ state. She prepares appropriate states over these according to
the above rules, as the protocol proceeds.  The joint state of both
parties together after the~$n$ commitment rounds then looks as given
below, for an arbitrary choice of~$b$ and the signs~$\set{s_{2j}}$
picked by Bob. Here, the qutrits sent by Alice to Bob are underlined.
The entangled qutrits and sign bits can be identified from the
context.  The first two lines correspond to the part of the state
where Alice commits to a bit~$x$ by sending~$\ket{x+1}$, before she
can identify which bit~$b$ Bob has picked. The next two lines have the
part of the state where Alice has not committed to any bit, and sees
a~$\ket{b+1}$ in one of the qutrits Bob sent. The last term is the
remaining part of the state. Note that Alice can differentiate between
these three parts by examining her entangled qutrits and the qutrits
Bob sent her.  The odd lines contain the portion of the state
constructed
by Alice, and the even lines contain the portion prepared by
Bob (and sent to Alice).
\begin{eqnarray}
& & 
  \sum_{x = 0,1}
  \sum_{j = 1}^{k-1}
    \left[ \Tensor_{l = 1}^{j-1}
      \sqrt{1 - \lambda_l}\, \ket{0\underline{0}}
    \right] 
    \tensor
    \sqrt{{\lambda_j}\over 2}\, \ket{x+1,\underline{x+1}}
    \tensor
    \left[
      \frac{1}{\sqrt2} (\ket{0}\ket{\underline{\psi(x,0)}} + 
                       \ket{1}\ket{\underline{\psi(x,1)}})
    \right]^{k-j} \nonumber \\
& & ~~~~~~~~~~~ \Tensor \left( \sqrt{1-\epsilon}\,\ket{0} \right)^{j-1}
    \tensor
    \left[
      \Tensor_{l = j}^{k-1} \ket{\psi(b,s_{2l})}
    \right] \nonumber \\
\label{eqn-commitstate}
&+& \sum_{j = 1}^{k-1}
    \left[
      \Tensor_{l = 1}^{j} \sqrt{1 - \lambda_l}\, \ket{0\underline{0}}
    \right] \tensor
    \left[
      \frac{1}{\sqrt{2}}\,
      (\ket{0}\ket{\underline{\psi(b,0)}} 
      + \ket{1}\ket{\underline{\psi(b,1)}})
    \right]^{k-j} \\
& & ~~~~~~~~~~~
    \Tensor \left( \sqrt{1-\epsilon}\,\ket{0} \right)^{j-1}
    \tensor
    \left[ (-1)^{s_{2j}} \sqrt{\epsilon}\,\ket{b+1} \right] 
    \tensor
    \left[
      \Tensor_{l = j + 1}^{k-1} \ket{\psi(b,s_{2l})}
    \right] \nonumber \\
&+& \left[
      \Tensor_{l = 1}^{k-1} \sqrt{1 - \lambda_l}\, \ket{0\underline{0}}
    \right]
    \tensor
    \left( \sqrt{1-\lambda_k}\, \ket{0\underline{0}}
           + \sqrt{{\lambda_k}\over 2}\, \ket{1\underline{1}}
           + \sqrt{{\lambda_k}\over 2}\, \ket{2\underline{2}}
    \right) \nonumber \\
& & ~~~~~~~~~~~
    \Tensor
    \left( \sqrt{1-\epsilon}\,\ket{0} \right)^{k-1}. \nonumber
\end{eqnarray}

Since~$n = 2k-1$ is odd, Bob reveals his bit~$b$ first. W.l.o.g., we
may assume that Alice would like to bias the coin towards~$0$.  She
therefore sends~$a = b$ in the part of her state where her entangled
qutrit is not equal to~$\bar{b}+1$ (which corresponds to a commitment
which she cannot change).  We will consider the residual state after
Alice has sent back~$k-i$ signs, in reverse order. This is the {\em
unnormalised\/} part of the state~(\ref{eqn-commitstate}) that has not
been rejected by Bob. We will prove by induction that Alice can
locally transform the residual state to a state~$\ket{\phi_i}$ after
every two rounds of sign exchange.  This state is similar in form to the
joint
state~(\ref{eqn-commitstate}) above, except that the first part is
projected onto the space where~$x = b$, and there is a factor
of~$\mu_i$ in the last term. The state~$\ket{\phi_i}$ is
displayed below:
\begin{eqnarray}
& & 
  \sum_{j = 1}^{i-1}
    \left[ \Tensor_{l = 1}^{j-1}
      \sqrt{1 - \lambda_l}\, \ket{0\underline{0}}
    \right] 
    \tensor
    \sqrt{{\lambda_j}\over 2}\, \ket{b+1,\underline{b+1}}
    \tensor
    \left[
      \frac{1}{\sqrt2} (\ket{0}\ket{\underline{\psi(b,0)}} + 
                       \ket{1}\ket{\underline{\psi(b,1)}})
    \right]^{i-j} \nonumber \\
& & ~~~~~~~~~~~ \Tensor \left( \sqrt{1-\epsilon}\,\ket{0} \right)^{j-1}
    \tensor
    \left[
      \Tensor_{l = j}^{i-1} \ket{\psi(b,s_{2l})}
    \right] \nonumber \\
\label{eqn-phi}
&+& \sum_{j = 1}^{i-1}
    \left[
      \Tensor_{l = 1}^{j} \sqrt{1 - \lambda_l}\, \ket{0\underline{0}}
    \right] \tensor
    \left[
      \frac{1}{\sqrt{2}}\,
      (\ket{0}\ket{\underline{\psi(b,0)}} 
      + \ket{1}\ket{\underline{\psi(b,1)}})
    \right]^{i-j} \\
& & ~~~~~~~~~~~
    \Tensor \left( \sqrt{1-\epsilon}\,\ket{0} \right)^{j-1}
    \tensor
    \left[ (-1)^{s_{2j}} \sqrt{\epsilon}\,\ket{b+1} \right] 
    \tensor
    \left[
      \Tensor_{l = j + 1}^{i-1} \ket{\psi(b,s_{2l})}
    \right] \nonumber \\
&+& \left[
      \Tensor_{l = 1}^{i-1} \sqrt{1 - \lambda_l}\, \ket{0\underline{0}}
    \right]
    \tensor
    \left( \sqrt{\mu_i(1-\lambda_i)}\, \ket{0\underline{0}}
           + \sqrt{{\lambda_i}\over 2}\, \ket{b+1,\underline{b+1}}
    \right) \nonumber \\
& & ~~~~~~~~~~~
    \Tensor
    \left( \sqrt{1-\epsilon}\,\ket{0} \right)^{i-1}, \nonumber
\end{eqnarray}
where the numbers~$\mu_i$ are as follows:
\begin{eqnarray}
\mu_k & = & 1 \nonumber \\
\label{eqn-mu}
\mu_{i-1} & = & (1-\epsilon)^2 \mu_i + \frac{\epsilon}{2}(3-\epsilon).
\end{eqnarray}
We can now also specify the parameters~$\lambda_i$:
\begin{eqnarray}
\label{eqn-lambda}
\lambda_i & = & \frac{\epsilon/2}{\mu_i (1-\epsilon) + \epsilon/2}.
\end{eqnarray}

Clearly, the state~$\ket{\phi_k}$, when none of the signs have been
revealed by Alice is of this form, with~$\mu_k = 1$.  Assume that this
is also the case for some~$i \le k$.  We will show by induction that
the state after Alice has revealed~$k-i+1$ sign bits~$s_{2k-1},
s_{2k-3}, \ldots, s_{2i-1}$ may be transformed to~(\ref{eqn-phi}) and
that equation~(\ref{eqn-mu}) holds. 

To send the sign~$s_{2i-1}$, Alice does the following. The sign in
part of the state in the first two summations of the
state~$\ket{\phi_i}$ in equation~(\ref{eqn-phi})) is ``pre-computed''
(in the sign qubit). To compute the sign in the last term, Alice first
sets~$b+1$ to~$1$ in the~$i$-th entangled qutrit, does a Hadamard
transform, and exchanges that entangled qubit with the~$i$-th sign
qubit. She then measures the~$i$-th sign qubit, and sends it across.
It is easily seen that the unnormalised state that remains after Bob
has checked the~$i$-th qutrit sent by Alice is as follows.\footnote{
  Actually, the state is a mixture of two states which are
  both~$1/\sqrt{2}$ times the state given. The mixture arises because
  of the two possible values of the sign bit Alice sends
  for~$s_{2i-1}$.  The mixture is of course equivalent to the single
  state shown.} 
Here, we have written the last terms of the first two summations in
equation~(\ref{eqn-phi}) separately in lines~5 and~7 to facilitate the
rest of the proof.
\begin{eqnarray*}
& & 
  \sum_{j = 1}^{i-2}
    \left[ \Tensor_{l = 1}^{j-1}
      \sqrt{1 - \lambda_l}\, \ket{0\underline{0}}
    \right] 
    \tensor
    \sqrt{{\lambda_j}\over 2}\, \ket{b+1,\underline{b+1}}
    \tensor
    \left[
      \frac{1}{\sqrt2} (\ket{0}\ket{\underline{\psi(b,0)}} + 
                       \ket{1}\ket{\underline{\psi(b,1)}})
    \right]^{i-1-j} \\
& & ~~~~~~~~~~~ \Tensor \left( \sqrt{1-\epsilon}\,\ket{0} \right)^{j-1}
    \tensor
    \left[
      \Tensor_{l = j}^{i-1} \ket{\psi(b,s_{2l})}
    \right] \\
&+& \sum_{j = 1}^{i-2}
    \left[
      \Tensor_{l = 1}^{j} \sqrt{1 - \lambda_l}\, \ket{0\underline{0}}
    \right] \tensor
    \left[
      \frac{1}{\sqrt{2}}\,
      (\ket{0}\ket{\underline{\psi(b,0)}} 
      + \ket{1}\ket{\underline{\psi(b,1)}})
    \right]^{i-1-j} \\
& & ~~~~~~~~~~~
    \Tensor \left( \sqrt{1-\epsilon}\,\ket{0} \right)^{j-1}
    \tensor
    \left[ (-1)^{s_{2j}} \sqrt{\epsilon}\,\ket{b+1} \right] 
    \tensor
    \left[
      \Tensor_{l = j + 1}^{i-1} \ket{\psi(b,s_{2l})}
    \right] \\
&+& \left[ \Tensor_{l = 1}^{i-2}
      \sqrt{1 - \lambda_l}\, \ket{0\underline{0}}
    \right] 
    \tensor
    \left(
      \sqrt{{\lambda_{i-1}}\over 2}\, \ket{b+1,\underline{b+1}}
    \right) \\
& & ~~~~~~~~~~~
    \Tensor \left( \sqrt{1-\epsilon}\,\ket{0} \right)^{i-2}
      \tensor \ket{\psi(b,s_{2(i-1)})} \\
&+& \left[
      \Tensor_{l = 1}^{i-2} \sqrt{1 - \lambda_l}\, \ket{0\underline{0}}
    \right] \tensor
    \left( \sqrt{1-\lambda_{i-1}}\, \ket{0\underline{0}} \right) \\
& & ~~~~~~~~~~~
    \Tensor \left( \sqrt{1-\epsilon}\,\ket{0} \right)^{i-2}
    \tensor
    \left[ (-1)^{s_{2(i-1)}} \sqrt{\epsilon}\,\ket{b+1} \right] \\
&+& \left[
      \Tensor_{l = 1}^{i-2} \sqrt{1 - \lambda_l}\, \ket{0\underline{0}}
    \right]
    \tensor
    \left( \sqrt{1-\lambda_{i-1}}\, \ket{0\underline{0}} \right) \\
& & ~~~~~~~~~~~
    \Tensor
    \left( \sqrt{1-\epsilon}\,\ket{0} \right)^{i-2}
    \tensor \sqrt{1-\epsilon}\,\ket{0}
      (\mu_i (1-\epsilon) + \epsilon/2)^{1/2}
\end{eqnarray*}
Now, when Bob sends the sign~$s_{2(i-1)}$ to Alice, she rotates
the~$(i-1)$-th qutrit that Bob sent her in all but the last two terms
in the sum above, to~$\ket{0}$. She also rotates that qutrit in the
last two terms from
$$
(\mu_i (1-\epsilon) + \epsilon/2)^{1/2} \sqrt{1-\epsilon}\; \ket{0} ~+~
(-1)^{s_{2(i-1)}} \sqrt{\epsilon}\; \ket{b+1}
$$
to~$\sqrt{\mu_{i-1}} \ket{0}$, whereby~$\mu_{i-1} = \mu_i (1-\epsilon)^2
+
\epsilon(1-\epsilon)/2 + \epsilon$. This proves the induction step.

At the final round of the protocol~$\pee_n$, the state that they are
left with is 
$$
\sqrt{\mu_1(1 - \lambda_1)}\, \ket{0\underline{0}}
~+~ \sqrt{\frac{\lambda_1}{2}}\, \ket{b+1,\underline{b+1}}.  
$$
Following Alice's strategy for computing the sign as above, we see
that the probability with which Alice succeeds in passing Bob's checks
is (using equation~(\ref{eqn-lambda}))
\begin{eqnarray}
\label{eqn-prob}
\left(
\sqrt{\mu_1(1 - \lambda_1)(1-\epsilon)}
+ \sqrt{\lambda_1 \epsilon/2} \right)^2
& = & \mu_1 (1-\epsilon) + \epsilon/2.
\end{eqnarray}

Solving the recurrence for~$\mu_i$ given in equation~(\ref{eqn-mu}),
we get 
$$
\mu_i ~~=~~ (1-\epsilon)^{2(k-i)}
            + \frac{(3-\epsilon)}{2(2-\epsilon)}
              (1 - (1-\epsilon)^{2(k-i)}),
$$
and so that from equation~(\ref{eqn-prob}),
\begin{eqnarray}
\textrm{For } n \textrm{ odd,~~~} A_n(\epsilon)
  & \ge & \frac{\epsilon}{2} + (1-\epsilon)^n
    + \frac{(3-\epsilon)(1-\epsilon)}{2(2-\epsilon)}
      (1 - (1-\epsilon)^{n-1}) \nonumber \\
\label{eqn-an1}
  & =   & \frac{1}{2(2-\epsilon)} ( 3-2\epsilon + (1-\epsilon)^{n+1} ).
\end{eqnarray}

The analysis in the case that~$n = 2k$ is even is similar, 
except for the rule
Alice uses to compute the bit she sends to Bob 
(since she is supposed to reveal
her bit before Bob reveals his bit). In this case,
she sends the bit~$x$ if any
of her entangled qutrits is in state~$\ket{x+1}$. Else,
if any of Bob's qutrits is in
state~$b+1$, she sends~$b$. In the remaining case,
she sends~$0,1$ with equal
amplitude~$1/\sqrt{2}$. This leads to a state similar 
to~$\ket{\phi_k}$ above after Bob reveals his bit, and the last sign he
used,
except that here~$\mu_k = (1+\epsilon)/2$. So,
we get:
\begin{eqnarray}
\textrm{For } n \textrm{ even,~~~} A_n(\epsilon)
  & \ge & \frac{\epsilon}{2} + (1-\epsilon)^{n-1} (1+\epsilon)/2
          + \frac{(3-\epsilon)(1-\epsilon)}{2(2-\epsilon)}
            (1 - (1-\epsilon)^{n-2)}) \nonumber \\
\label{eqn-an2}
  & =   & \frac{1}{2(2-\epsilon)} ( 3-2\epsilon - (1-\epsilon)^{n+1} ).
\end{eqnarray}

From equations~(\ref{eqn-an1}), (\ref{eqn-an2}), and~(\ref{eqn-bn0}),
we can deduce lower bounds for~$B_n(\epsilon)$ as well, for~$n \ge 2$.
This expression matches the one for~$B_1(\epsilon)$. Thus,
\begin{eqnarray}
\label{eqn-bn}
\textrm{For all } n,~~~ B_n(\epsilon)
  & \ge & \frac{1}{2(2-\epsilon)}
          ( 3 - \epsilon + (-1)^n (1-\epsilon)^{n+1} ).
\end{eqnarray}

To determine the bias achieved, we examine the maximum of the cheating
probabilities attained by Alice and Bob.  Note that~$A_n(\epsilon) +
B_n(\epsilon) \ge 3/2$ for all~$n,\epsilon$. Thus, the bias of the
protocol~$\pee_n$ is at least~$3/4$ for any~$n$ and~$\epsilon$. Since
we would like this bias to be as small as possible, we optimise the
maximum cheating probability with respect to~$\epsilon$.

For odd~$n$, $A_n(0) = B_n(1) = 1$, and~$A_n(1) = B(0) = 1/2$,
$A_n$ is monotonically decreasing, and $B_n$ is monotonically
increasing with respect to~$\epsilon \in  [0,1]$. Thus, the maximum
bias achievable is minimised when~$A_n(\epsilon) = B_n(\epsilon)$. This
condition is satisfied when~$\epsilon = \epsilon_0$ such that
\begin{eqnarray}
\label{eqn-eps1}
(1-\epsilon_0)^{n+1} & = & \frac{\epsilon_0}{2} \textrm{ ~~~~that is,}
\\
\epsilon_0 & = & \frac{1}{n+1}(\ln n - \ln\ln n + \Theta(1)). \nonumber
\end{eqnarray}
Using equation~(\ref{eqn-eps1}) we can verify that the lower bound
on~$A_n(\epsilon_0)$ is~$3/4$.

For even~$n$, $A_n(0) = A_n(1) = 1/2$, and~$B(0) = B_n(1) = 1$,
$A_n$ is concave, $B_n$ is convex, and~$B_n(\epsilon) \ge
A_n(\epsilon)$ for~$\epsilon \in [0,1]$. Thus, the bias achievable is
minimised when~$B_n$ is, i.e., for~$\epsilon = \epsilon_0$ such that
\begin{eqnarray}
\label{eqn-eps2}
(1-\epsilon_0)^n & = & \frac{1}{(2-\epsilon_0)n + 1}
                     \textrm{ ~~~~that is,} \\
\epsilon_0 & = & \frac{1}{n} \ln(2n - \ln n +\Theta(1)). \nonumber
\end{eqnarray}
The expression for the cheating probability~$B_n(\epsilon_0)$ then
evaluates to at least~$\frac{3}{4}$, as may be seen by using
equation~(\ref{eqn-eps2}).

This completes the analysis of the cheating strategies devised above.

\end{document}